# Electronic and structural properties of rhombohedral [111] and [110] oriented ultra-thin bismuth nanowires

Lida Ansari, Farzan Gity, and James C. Greer
Tyndall National Institute, Lee Maltings, Dyke Parade, Cork, T12 R5CP, Ireland

Structures and electronic properties of rhombohedral [111] and [110] bismuth nanowires are calculated with the use of density functional theory. The formation of an energy band gap from quantum confinement is studied and to improve estimates for the band gap the *GW* approximation is applied. The [111] oriented nanowires require surface bonds to be chemically saturated to avoid formation of metallic surface states whereas the surface of the [110] nanowires do not support metallic surface states. It is found that the onset of quantum confinement in the surface passivated [111] nanowires occurs at larger critical dimensions than for the [110] nanowires. For the [111] oriented nanowires it is predicted that a band gap of ~0.5 eV can be formed at a diameter of approximately 6 nm, whereas for the [110] oriented nanowires a diameter of approximately 3 nm is required to achieve a similar band gap energy. The *GW* correction is also applied to estimates of the electron affinity, ionisation potentials and work functions for both orientations of the nanowires for various diameters below 5 nm. The magnitude of the energy band gaps that arise in bismuth at critical dimensions of a few nanometers are of the same order as for conventional bulk semiconductors.

## 1. Introduction

Bismuth has long been studied due to its unusual structural, electronic and thermal properties [1,2,3]. The electronic structure in bulk form is semimetallic and displays a pronounced anisotropy in its electronic and thermal properties. The strong anisotropy is reflected in extremely small effective masses for the charge carriers and thermal conductivities along specific crystal directions. Related to the effective masses, electrons in bismuth can have unusually long mean free paths of a few hundred nanometers at room temperature ranging to the millimeter range at liquid helium temperatures [4]. The charge carriers in bismuth have a large Fermi wavelength, typically an order of magnitude larger than in most metals with reported values ranging between 12 and 70 nm [5,6,7]. The large Fermi wavelength of charge carriers results in the effects of quantum confinement being observable in two dimensional films [7,8] and nanowires (NWs) at relatively large spatial dimensions and sufficiently low temperatures [9,10,11,12]. In addition to strong quantum confinement effects in nanostructures, there are strong surface effects leading to surface states that decouple from the bulk and the surface may for certain orientations be treated as a two dimensional metal [7,8].

By reducing the dimensions of nanostructured bismuth to be on the order of the free charge carriers leads to the semimetallic character in bulk bismuth transitioning to semiconducting as an induced energy band gap is created as a result of quantum confinement. This leads to strikingly different behaviour in nanostructures of bismuth due to the semimetal-to-semiconductor transition [9,10,11,12]. Many properties for bismuth have been studied for thin films and NWs for spatial dimensions greater than 10 nanometers. Although quantum confinement effects can be seen in nanostructures at relatively large dimensions, to achieve a band gap that is significantly larger than the thermal energy $k_BT$ at room temperature requires confinement dimensions of less than 10 nm. For nanoelectronics applications, an energy band gap several times larger than $k_BT$ is required. It is known that below 6 nm commonly used semiconductors such as silicon and germanium become wide gap insulators. The "negative" band gap of bulk bismuth leads to a relatively smaller confinement induced band gaps relative to Si and Ge for similar confinement lengths, but the induced gap for Bi, as will be shown, can be of a similar magnitude to that of bulk silicon or germanium. Being able to achieve a 'conventional' semiconductor band gap at dimensions of a few nanometers along with unusal electronic properties, leads to nanostructured bismuth being of particular interest as a potential material for upcoming generations of electronic devices. The quantum confinement effect is more pronounced in NWs relative to the thin films due to charge carriers being constrained in two spatial dimensions rather than one, but the effects of band folding, surface faceting, and thermodynamic stability become much more complex for nanowires than for thin films. The semimetal-to-semiconductor transition in bismuth nanowires (BiNWs) has been studied with particle-in-a-box confinement models and a two state Lax model using effective masses from bismuth bulk [9,10]. A transition to an induced band gap of a few milli-electron volts is predicted to occur in range 35 to 70 nm with a strong dependence on both NW crystal orientation and shape. As noted in ref. [9,10], a two state model becomes increasingly less accurate below approximately 10 nm as the interaction between multiple bands becomes important. Density functional theory (DFT) studies of bismuth atomic chains [13] and for rhombohedral [110] oriented BiNWs with various cross sections [14] have been previously reported. However, for the purposes of nanoelectronics applications and understanding of ultra-scaled BiNWs, the magnitude of the band gap is a key physical quantity and unfortunately the prediction of band gap energies is a property for which approximate exchange-correlational functionals used in DFT calculations typically fail. To extend previous work, *ab initio* techniques are applied to study BiNWs with confinement dimensions below 5 nm and the *GW* approximation is applied to improve estimates for the band gap energies which is essential for correlating electronic structure and electrical properties to experiments.

Bismuth is a group V element and has a rhombohedral (arsenic) crystal structure. Along the rhombohedral [111] orientation or $C_3$ symmetry axis, the structure can be thought of as stacked bilayers whereby the three nearest neighbour bonds are within a single bilayer. Bonding between bilayers is weaker than within a bilayer, resulting in a two dimensional stacked structure. Thin films with the growth axis in the [111] direction have been prepared and studied using angle resolved photoemission spectroscopy (ARPES) [8] and the molecular beam epitaxy of bismuth films with a [111] growth direction is well understood [15]. Using top down fabrication techniques, it becomes possible to pattern vertical nanowires in the [111] films although there remains substantial challenges to development of lithography and etching techniques for the semimetals. An alternative approach to nanowire formation is injection of bismuth from a liquid melt into an anodic alumina template resulting in highly crystalline nanowires with a broad distribution of diameters [16]. X-ray diffraction (XRD) and selected area electron diffraction (SAED) patterns reveal that those nanowires with diameters greater than 50 nm have a preferred growth orientation of [100], whereas nanowire formed with this technique possessing diameters below 50 nm have a preferred growth orientation of [110] [16]. Free standing BiNWs can be experimentally realised. The anodic alumina template is dissolved following the injection of molten Bi. BiNWs with diameters between 10 nm to 200 nm and lengths on the order of 50 μm have been synthesized with this technique. Alternatively, top-down approaches can be employed to fabricate BiNWs through patterning of resist by e-beam lithography (EBL) on Bi film followed by etching the Bi using the patterned resist as a mask. Bi films have been grown by molecular beam epitaxy (MBE) on Si [111] substrates with a [111] bismuth film orientation parallel to the growth axis. Vertical BiNWs along the [111] orientation can be fabricated by etching thick Bi films although a loss of resolution for EBL as a result of the requirement to use thicker resist mask imposes limitations on the minimum diameter that could be achieved. For horizontal nanowires, the in-plane orientation of the Bi film defines the orientation of the BiNW. Ultra-thin Bi films allow thinner EBL resist masks and hence patterning of narrower nanowires. However, the crystal orientation of MBE grown Bi thin films is dependent on the thickness of the film and in order to circumvent the ambiguity of the in-plane orientation of the BiNW in the top-down approach, various nanowires patterned angled with respect to the substrate must first be fabricated to experimentally identify the preferred orientation of the BiNWs or other techniques aligning the in-plane crystal orientation to the substrate must be devised.

To investigate the electronic structure and properties of nanowires on length scales relevant for use in room temperature nanoelectronics implies that the confinement dimensions for bismuth should be of the order of 5 nm or less. DFT calculations for [111] and [110] nanowire orientations are performed to generate strain-free nanowires with the two chosen orientations being leading candidates for either top down or bottom up fabrication of BiNWs, respectively. For the length scales of less than 5 nm considered, the magnitude of the confinement induced band gaps are large enough that the interactions between conduction and valence bands can be ignored as a first approximation. This allows the Kohn-Sham eigenvalues to be interpreted as corresponding to the electronic band structures. The direct or non-direct nature of the band gaps and the effective masses at the conduction band minimum and valence band maximum are reported. As well, the surface states for pristine (*i.e.* without saturating the surface bonding with hydrogen or other terminating species) are considered; where metallic surface states are supported, their passivation through hydrogen termination is explored. A key physical quantity for nanoelectronics is the energy band gap, and as mentioned, approximate exchange-correlational functionals as commonly used in DFT calculations significantly underestimate band gaps in semiconductors. To provide a better representation of band gap energies and the related quantities of ionisation potentials and electron affinities, the *GW* approximation is applied. The *GW* approximation has been shown to provide significantly improved band gap estimates for semiconductors and insulators [17]. The *GW* corrected band structures allow for electron affinities and band gaps to be used for estimating band alignments between two different NW orientations and/or different diameters or more specifically, for differing cross sections.

## 2. Structural models and computational methods

The [111] oriented BiNWs are constructed with hexagonal cross sections with 0.9, 1.5, 1.9, 2.6, 3.2 and 4.3 nm lengths along the $C_3$ rotation symmetry axe and for the [110] oriented nanowires approximately square cross section NW with lengths along the diagonal of 1.0, 1.6, 2.3, 3.0 and 4.3 nm are constructed. For simplicity in the following, these lengths will be subsequently referred to as the NW 'diameter' as is convention in the literature. Representative structures for both orientations are shown in figure 1. The structures are repeated periodically along the nanowire axis resulting in single crystal nanowires with infinite lengths.

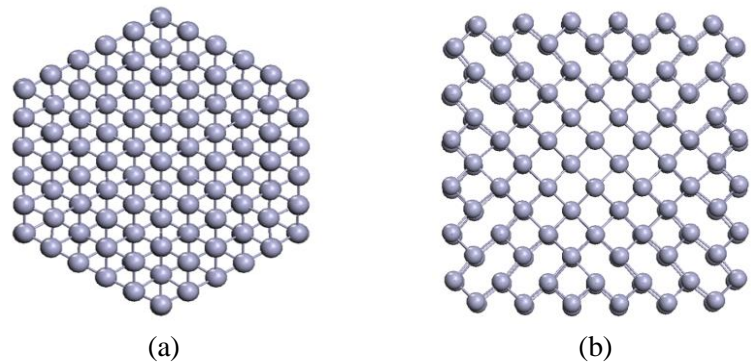

(a)          (b)

**Figure 1.** Cross sectional views of representative bismuth nanowires used in these studies. (a) Bismuth nanowire with rhombohedral [111] orientation. Note the structure is constructed from bulk bismuth and not relaxed. Hydrogen termination not shown. (b) Bismuth nanowire with rhombohedral [110] orientation. The structure is stable without surface passivation and the relaxed structure is shown.

The electronic energy for is calculated using DFT as implemented in OpenMX [18] and Quantum Espresso [19]. OpenMX and Quantum Espresso calculate the electronic energy within the framework of Kohn-Sham DFT using pseudo-atomic localized basis and plane wave basis functions, respectively. The local density approximation (LDA) for the exchange correlation potential and norm conserving pseudopotentials are used for both the atomic and plane wave basis sets. A numerical atomic orbital (NAO) basis sets s3p3d2 is used for Bi and for the hydrogen terminations a s2p2 basis is used. The plane wave calculations are performed with an energy cutoff of 60 Rydberg. Structural optimisation is achieved by minimisation of the total energy with respect to the atomic positions until the maximum force component becomes less than 0.01 eV/Å. Periodic boundary conditions are applied creating simulation supercells and the cell dimensions transverse to the nanowire long axis are chosen to be greater than 2 nm to neglect the interaction between the periodic images of neighbouring nanowires. The localised basis sets enable the larger diameter nanowire to be studied with reduced computational demand and the numerical pseudo-atomic orbitals basis sets are confirmed by comparing to plane wave (or approximately complete basis set) calculations for the smaller nanowires. All nanowires were computed using the NAO basis sets and the plane wave calculations are repeated for diameters less than 2 nm for the [111] orientation and less than 2.3 nm diameter for the [110] orientation. In addition to assessing the completeness of the basis sets for the NAO calculations, the plane wave DFT/LDA calculations form the starting point for the *GW* calculations.

To provide an improved quantitative description of electronic structure, the *GW* approximation (*G*: Green's function; *W*: screened Coulomb interaction) method is used to improve excitation energies that compare well with experiment for a wide range of semiconductors and insulators [17]. Green's function theory provides quasiparticle energies using the electron self-energy operator containing the effects of exchange and correlations with other quasiparticles. The Yambo program is used in this work to calculate the *GW* approximation correction to quasiparticle energies within first order perturbation theory starting from the Kohn-Sham eigenstates [20]. Due to the computational demands for *GW* calculations in comparison with DFT calculations, this approach is used to calculate [111] nanowires of diamter 0.9, 1.5 and 1.9 nm, and [110] nanowires of diamter 1.0, 1.6 and 2.3 nm. The *GW* calculations are used to 'calibrate' the confinement induced band gap at the smaller NW diameters and this correction is extrapolated to correct the band gap energies for the NWs with larger diameters. For the small cross section BiNWs considered in this study, the quantum confinement induced gap results in semiconducting behaviour with the dielectric response function displaying a single sharp peak in the energy range of interest allowing the plasmon pole approximation to be applied. The effective Coulomb interaction is set to zero in real space within the vaccum region such that quasi-particles do not interact with their periodic images [21]. As a result, the periodicity of the structure is reduced using cutoff Coulomb interaction in the direction transverse to the NWs. In the *GW* calculations the method in ref. [22] has been used enabling for improved convergence in the band gap energies with a reduced set of empty states as well as requiring the *GW* correction k-points of interest. A $16 \times 1 \times 1$ k-point Monkhorst-Pack grid in the irreducible Brillouin zone is used with all G-vectors have being used in the *GW* calculations.

## 3. Electronic structure and properties

Bulk bismuth is a semimetal. In terms of its electronic structure, it may be thought of as having a conduction band minimum (CBM) at the L symmetry point and a valence band minimum (VBM) at the T symmetry point in the Brillouin zone. At both the L and T points there are direct band gaps, the band gap at the L point is approximately 40 meV at a temperature of 300 K. A range of values have been reported for the direct band gap at the T point ranging from 50 to 400 meV at 300 K. However, the defining feature making bismuth a semimetal is that the CBM at the L point is approximately 60 to 100 meV lower in energy at room temperature relative to the VBM at the T point. This indirect band overlap between the CBM and VBM gives rise to bulk bismuth's semimetallic properties. The shape of the energy minima about the L and T points is highly anisotropic and for energies relevant for room temperature charge transport, the energy surface about the CBM at the L point is highly nonparabolic. Although the strong nonparabolicity of the bands requires corrections to the effective mass model, nonetheless extremely small effective masses have been deduced for bismuth from mobility measurements ranging from 0.001 $m_0$ to 0.26 $m_0$ where $m_0$ is the free electron mass. The range of values reflects the strong anisotropy in the band structure and for the smallest values, accounts for the extremely long charge carrier wavelengths leading to pronounced quantum confinement effects at relatively large nanostructure dimensions. Additionally, it is known that surfaces of bismuth can support metallic states that are only weakly coupled to the bulk bismuth states [7]. As can be anticipated by the novelty of the electronic structure for bulk bismuth and bismuth surfaces, the electronic structure of bismuth nanowires (NWs) can be expected to be complex and to display new properties.

First consider the [111] oriented nanowires. As the atom positions are allowed to relax in the DFT structure optimisations, it is found that the NWs as constructed become unstable which is reflected in considerable atomic rearrangements and a loss in crystallinity. However, introducing hydrogen termination to the surface atoms in the [111] nanowires, the structures become stable and retain the bonding motif of bulk bismuth with either one or two hydrogen atoms bonded to the Bi atoms leading to threefold coordinated Bi. Fig. 2 shows the resulting density of states for the [111] BiNW with diameter of 3.1 nm with and without surface termination. As the unterminated NW is unstable, the geometry of the nanowire is relaxed with hydrogen termination bonding to surface Bi atoms. The electronic structure of the nanowire is calculated first at the relaxed geometry and repeated with the hydrogen surface terminations removed. The DoS from these two calculations is shown in fig. 2. Without the surface terminating hydrogen atoms, the nanowire is seen to be strongly metallic. When the NW surface is saturated by bonding to hydrogen atoms, a band gap on the order of 0.4 eV is seen to emerge at the level of the DFT/LDA calculations. Conversely, the [110] oriented BiNWs are found to be stable and display a band gap for all NW cross sections studied. Relatively small surface relaxations allow for the surface atoms to achieve equilibrium positions that represent small displacements relative to ideal bulk positions, and the formation of a band gap due to quantum confinement with the absence of surface terminating groups indicates there are no metallic surface states.

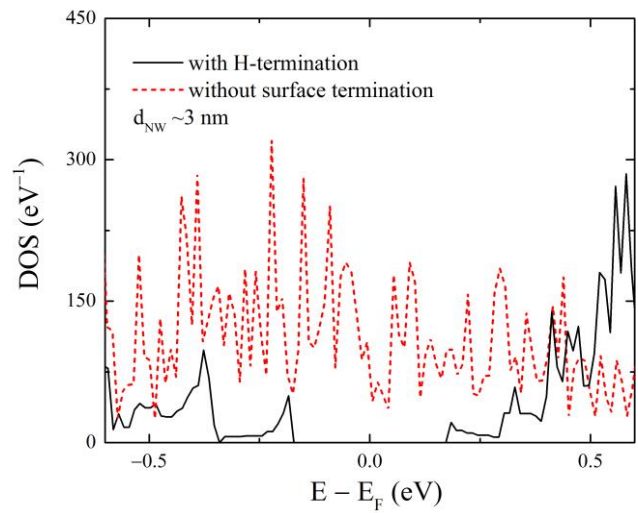

**Figure 2**. Density of states (DoS) for a [111] oriented bismuth nanowire with diameter of 3.1 nm with and without hydrogen termination.

The band structure of BiNWs with various diameters along the [111] and [110] orientations obtained from the Kohn-Sham eigenvalues arising from the DFT/LDA calculations are shown in fig. 3 for the two NW orientations and for various cross sections. As can be seen from the band structure, the band gap energy increases with decreasing nanowire diameter consistent with the anticipated effects of reduced dimensionality. The DFT/LDA results confirms the prediction that BiNWs trigonal orientation results in a larger value of band gap for comparable confinement lengths relative to the [110] oriented NWs. The emergence of a band gap for the [111] oriented NW for larger confinement dimensions is consistent with the semi-empirical confinement model prediction that [111] oriented NWs should display larger confinement relative to other orientations [10,16]. The semi-empirical models rely on the effective mass parameters from bulk Bi, which as mentioned is not believed to be applicable for nanowires with confinement dimensions of a few nanometers [9,10] and this expectation is reinforced by the DFT/LDA band structures. Nonetheless, the calculations suggest the band gap widening for the [111] orientation remains more pronounced even for BiNW with less than 5 nm diameters.

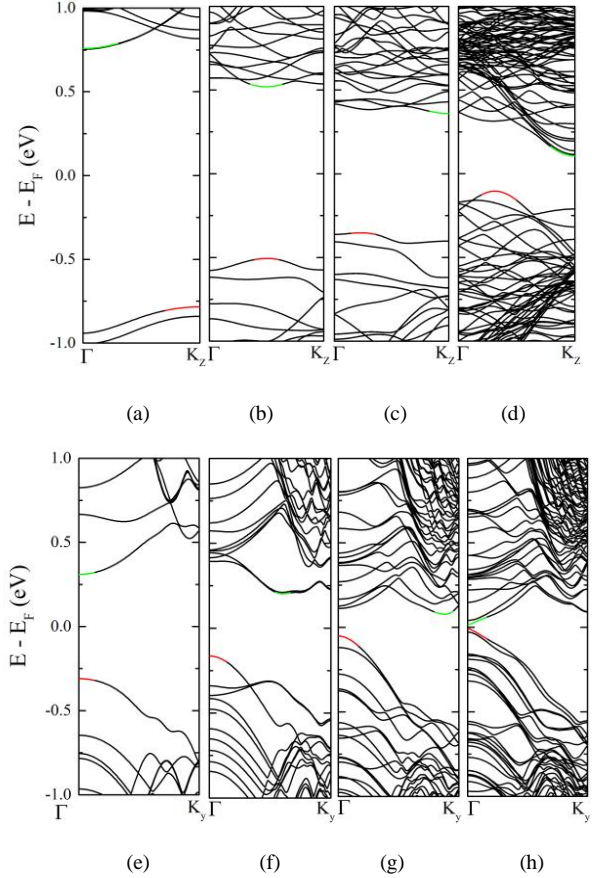

**Figure 3**. Band structure of BiNWs with [111] orientation along the NW long axis with diameters of (a) 1.2 nm, (b) 1.6 nm, (c) 1.9 nm, (d) 4.2 nm, and with [110] orientation along the NW long axis with diameters of (e) 1 nm, (f) 1.6 nm, (g) 2.3 nm, and (h) 3 nm. Energies are referenced to the Fermi level and the green highlighted regions show the conduction band minima and the red highlighted regions indicate the valence band maxima.

For the largest [111] oriented BiNW displayed in fig. 3(d) with critical dimensions of ~4.2 nm, the overlap between the conduction and valence bands is lifted due to quantum confinement and the band folding is such that an indirect band gap material is formed. The critical dimension in fig. 3(c) is ~2 nm and substantial modification to the electronic structure relative to the 4.2 nm BiNW is seen. The band gap increases, the conduction and valence bands are flatter and the band gap to within the accuracy of the calculations is indirect. Reducing the critical dimension to ~1.6 nm results in a further increase in the band gap and with a slightly indirect nature. For the BiNW with the smallest cross section considered for the [111] orientation, as expected, results a further increase in the band gap energy and the valence band maximum (VBM) is located at the Brillouin zone edge and the conduction band minimum (CBM) is located at Γ. This in contrast to the larger BiNW with critical dimensions of ~4.2 nm which is also of a strong indirect nature but with the CBM at the Brillouin zone edge and the VBM between the Γ point and the Brillouin zone edge. The quantum confinement effect is significantly less in the [110] oriented NWs with a small direct band gap at the Γ point of the order of $k_B T$ at room temperature emerging at the critical confinement dimension of ~3 nm increasing at the Γ point to the order of several hundred milli-electron volts for confinement dimensions of approximately 2.3, 1.6 and 1 nm. The lowest energy conduction states at the edge of the Brillouin zone do not increase in energy as quickly as the lowest energy conduction states at Γ. Hence the CBM for the 2.3 nm BiNW is for the states close to the edge of the Brillouin zone, for the 1.6 nm wire the CBM is between the Γ point and the Brillouin zone edge, whereas a direct gap is seen for the smallest [110] NW. Unlike the [111] NWs, for the [110] oriented NWs the VBM is at the Γ point for all cross sections studied.

The calculated band gaps are shown as a function of confinement critical dimension in fig. 4. As anticipated, the DFT/LDA band gaps calculated from the Kohn-Sham eigenvalues provide significantly lower band gap energy relative to the *GW* corrected eigenvalues. As the *GW* calculations have been shown to provide better band gap estimates across a wide range of materials [17], attention is focused on the *GW* predictions. Simple models of quantum confinement predict a power law behaviour for the energy band gap in analogy to the behaviour of confined states in the particle-in-a-box model. However, the simple confinement model describes the evolution of a single state as the system size is reduced. As shown in fig. 3, the band gap energy arises from different band extrema at different energies. Hence, the band gap energies vary in a nontrivial way with the critical confinement dimension. Nonetheless, the band gap variations can be well described by a power law fitting function that interpolates between the *GW* calculated band gaps and experimental values obtained for larger nanowire diameters [23]. The solid black curve in fig. 4 gives the model finding from ref. [24] where it is seen that the band gaps are overestimated which, as will be seen, can be ascribed to the larger effective masses due to band folding in these small nanostructures than that assumed in the model calculations.

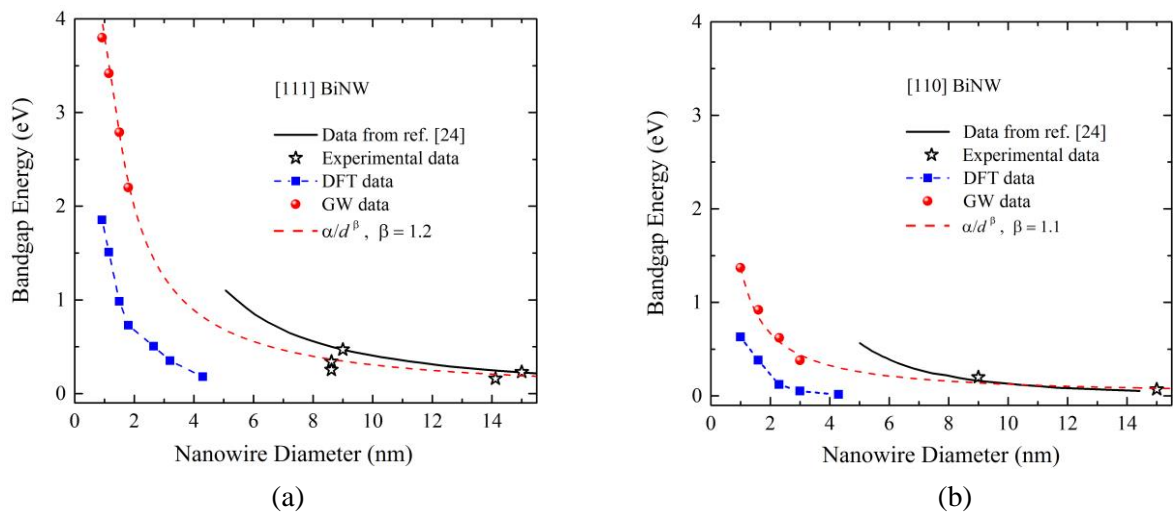

**Figure 4.** Bandgap modulation of BiNW along (a) [111] orientation, and (b) [110] orientation using DFT and implementing *GW* corrections by nanowire diameter. The solid curve is taken from ref. [24] where the bandgap value of BiNWs extracted through numerical model based on bulk effective mass of Bi. Data points labeled ☆ are taken from experimental data given in ref. [23]. Dashed curves indicate fitting of *GW* energies and experimental energies to a power law $\alpha/d^\beta$ with $d$ the NW diameter and fitting to the power law allows estimates of the band gap energy for a range of nanowire diameters.

For both orientations [111] and [110] the band structure is complex and varies significantly for relatively modest changes to the nanowire diameter. This is also reflected in the non-monotonic behaviour of the effective masses as a function of NW diameter. Usually several bands are required to adequately describe charge transport but in the BiNWs care will be required in determining effective masses and the ordering of energy bands for varying NW diameter. Effective masses calculated for the conduction and valence band extrema highlighted in fig. 3 for [111] and [110] oriented BiNWs are reported in table 1.

**Table 1**. Effective masses as calculated for rhombohedral [111] and [110] oriented BiNWs. The critical dimension corresponds to the NW diameter as defined in the text and given in nanometer. $m_e$ is the electron effective mass and $m_h$ is the hole effective mass in units of the free electron mass $m_0$.

| BiNW orientation | Critical dimension (nm) | $m_e / m_0$ | $m_h / m_0$ |
|---|---|---|---|
| [111] | 1.2 | 0.78 | 1.17 |
| | 1.6 | 0.37 | 0.45 |
| | 2.0 | 0.38 | 0.46 |
| | 3.2 | 0.89 | 1.30 |
| | 4.2 | 0.22 | 0.18 |
| [110] | 1.0 | 1.51 | 1.63 |
| | 1.6 | 0.45 | 1.20 |
| | 2.3 | 0.30 | 0.90 |
| | 3.0 | 0.22 | 0.70 |
| | 3.6 | 0.16 | 0.55 |

The work function $W_F$ is the energy required to removing an electron from the Fermi level to the vacuum level ($W_F = E_{vac} - E_F$). Knowledge of the work function difference between two materials aids in prediction of band alignments and potential barrier heights at interfaces, likewise in the context of NWs, for differing diameters and potentially differing orientations and surface terminations. Fermi level pinning can overshadow the effects of work function mismatch between two materials but nonetheless for junctions without a large density of states associated directly with the interface, the $W_F$ difference is a powerful diagnostic tool. Janak's theorem identifies the electron highest occupied orbital (HOMO) to the first electron ionisation potential in DFT playing a similar role to Koopmans' theorem in wave function theory [25,26], although the conclusion that the HOMO level corresponds to the ionisation potential has been questioned [27,28]. Nonetheless, it has been demonstrated in ref. [29] that Janak's theorem hold at least as an empirical observation for approximate DFT in describing metallic systems. An additional complication in calculation of accurate estimates for the ionisation energy is that localized atomic basis sets tend to underestimate the Kohn-Sham energies of surface states leading to too small of an ionisation potential. To improve the estimate for Kohn-Sham states with evanescent decay into the vacuum region, 'ghost' atoms are introduce in the vacuum region around the perimeter of the NW. Ghost atoms provide additional flexibility to atom centered basis sets through the introduction of zero pseudopotential cores without introducing additional electrons corresponding to the 'dummy' atom sites. The additional variational freedom provided by the ghost atoms allows for improved convergence toward the full basis set limits for the Kohn-Sham eigenvalue calculations. The work functions for BiNWs with diameters up to 2 nm were calculated using the localized basis sets augmented as described and compared to the plane-wave DFT calculations yielding good agreement for the HOMO and LUMO Kohn-Sham eigenvalues. This is taken to indicate that the augmented localized basis is of comparable quality to the plane-wave basis allowing its use for larger diameter NWs with less computational time required relative to the corresponding plane-wave calculations. Using this approach, the vacuum energy estimated by interpreting the Kohn-Sham HOMO energy as the VBM and ionisation potential and the Kohn-Sham lowest unoccupied orbital (LUMO) corresponds to the CBM. The work function is estimated as the energy between the vacuum level and the Fermi energy, similarly the electron affinity ($E_A$) is estimated as the difference between the vacuum level and the CBM ($E_A = E_{vac} - E_C$). However as discussed in the preceding section in the context of band gap energy estimates, it is known that the Kohn-Sham eigenvalues do not accurately describe the electron ionisation potential and electron affinity. Hence these levels are corrected using the *GW* approximation. Fig. 5 shows the calculated band diagram for the BiNWs. Fig. 5(a) and 5(b) shows the DFT/LDA and *GW* corrected energy levels, respectively for the [111] oriented NWs and fig. 5(c) and 5(d) shows the DFT/LDA and *GW* corrected energy levels, respectively for the [110] oriented NWs.

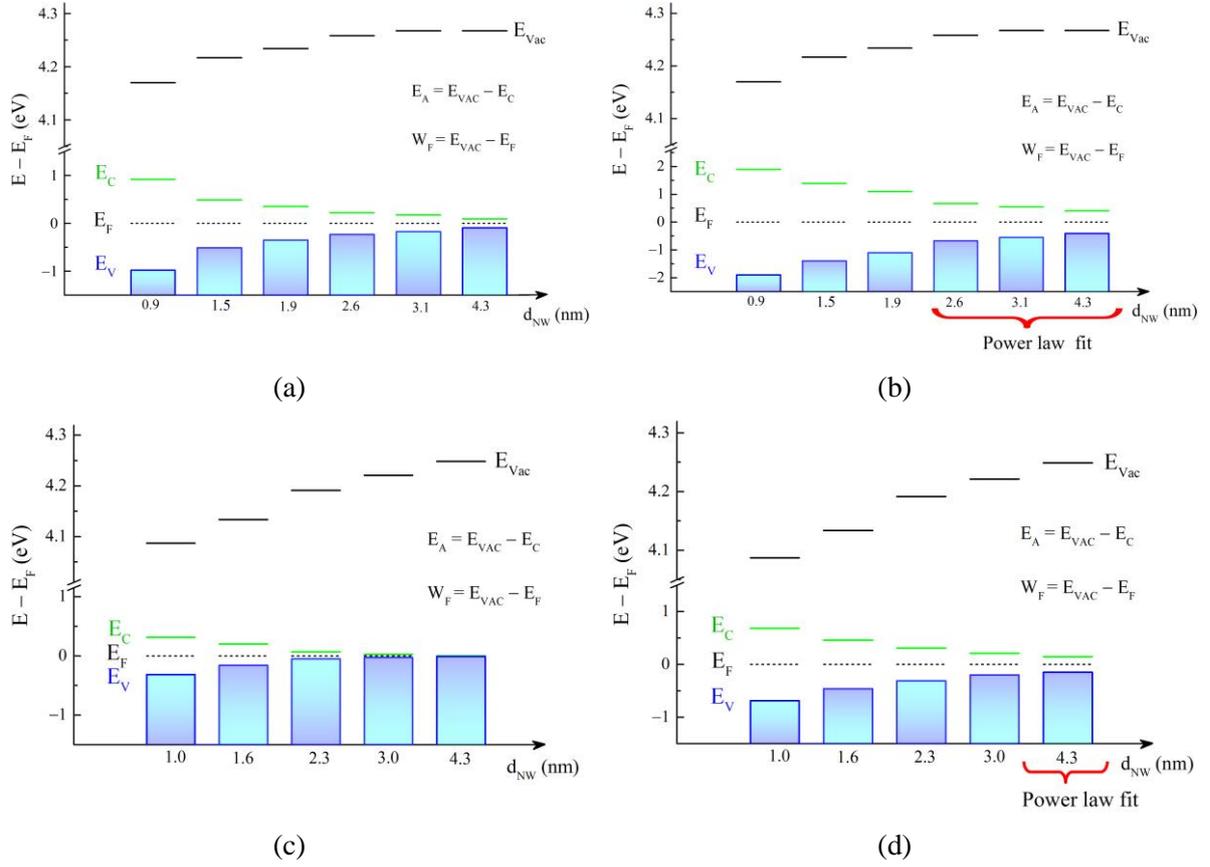

**Figure 5.** Energy band diagram of BiNWs along [111] orientation (a) using DFT, and (b) after implementing *GW* corrections as functions of nanowire diameter. Similarly, for [110] (c) using DFT and (d) after implementing *GW* corrections. For the NW diameters labeled "Power law fit", the *GW* correction is estimated by interpolation between the explicit *GW* calculations for smaller diameter NWs and measured values for larger diameter NWs.

The band gaps are larger for the [111] oriented NWs as seen in different representations in figs. 3, 4 and 5 at larger NW critical dimensions relative to the [110] oriented NWs. However, fig. 5(b) and 5(d) reveal comparable values of the work function for both orientations. When forming a junction between a narrow BiNW and a metallic contact, the Fermi levels between the two regions will align. The amount of charge transfer to align the levels dictated by the difference in the work functions between the two material regions; this behaviour is expected to be similar for the two orientations considered. The possibility exists to fabricate a semimetal NW with a diameter larger than the critical dimension required for the onset of quantum confinement, or for room temperature applications for critical dimensions such that the induced band gap is not large in comparison to $k_BT \approx 26$ meV. If a region is subsequently thinned to critical dimensions below the semimetal-to-semiconducting transition, a Schottky barrier within a monomaterial semimetal NW can be formed [30]. Similar junctions are anticipated to be formed between a quantum confined semiconducting region and larger semimetallic region with either [111] and [110] orientated NWs, but with larger band offsets expected for the [111] BiNW due to the larger confinement induced energy gap.

As a final observation, the contour plots of the charge density for specific states averaged along the axial NW direction within a single unit cell at the VBM and the CBM for NW diameters $d_{NW}=$ 1 nm and 3 nm are compared in fig. 6 for the two orientations [111] and [110]. For the 1 nm [111] oriented BiNW, the VBM can be described as tending to be localised at the surface of the NW whereas the CBM state tends to the centre of the NW. There is not as clear of a distinction for the case of the 1 nm [110] oriented BiNW, although it can be said there is a node at the centre of the NW for the VBM whereas there is a charge accumulation at the centre of the NW for the CBM. However, the ~1 nm diameter of the NWs is such that a distinction between 'core' and 'surface' is not straightforward. For the larger 3 nm BiNWs, the distinction as to how the states are distributed becomes clearer. For the 3 nm [111] oriented BiNW, the charge distribution for the VBM appears as a first approximation to be uniformly distributed across the NW cross section, whereas for the CBM the state tends to be preferentially located toward the centre. Perhaps the most striking contrast is for the 3 nm [110] oriented BiNW. The VBM state is preferentially located toward the centre of the NW, whereas the CBM state can be identified as preferentially localised toward the NW surface. The nature of, for example, the CBM at the same NW cross section but differing orientations is fundamentally different for the [111] core state versus the [110] surface state. Additionally from the electronic band structure calculations, it is seen that the VBM and CBM can vary significantly with respect to BiNW diameter due to the effects of band folding and different quantum confinement effects for band minima of different effective masses. Hence although the discussion has focused on the different nature of edge states for NWs of the same diameter but of different orientations, due to the complex behaviour of the band structures with varying NW diameter, the edge states for BiNWs of the same orientation but differing diameters can also be anticipated to display similarly varied characteristics.

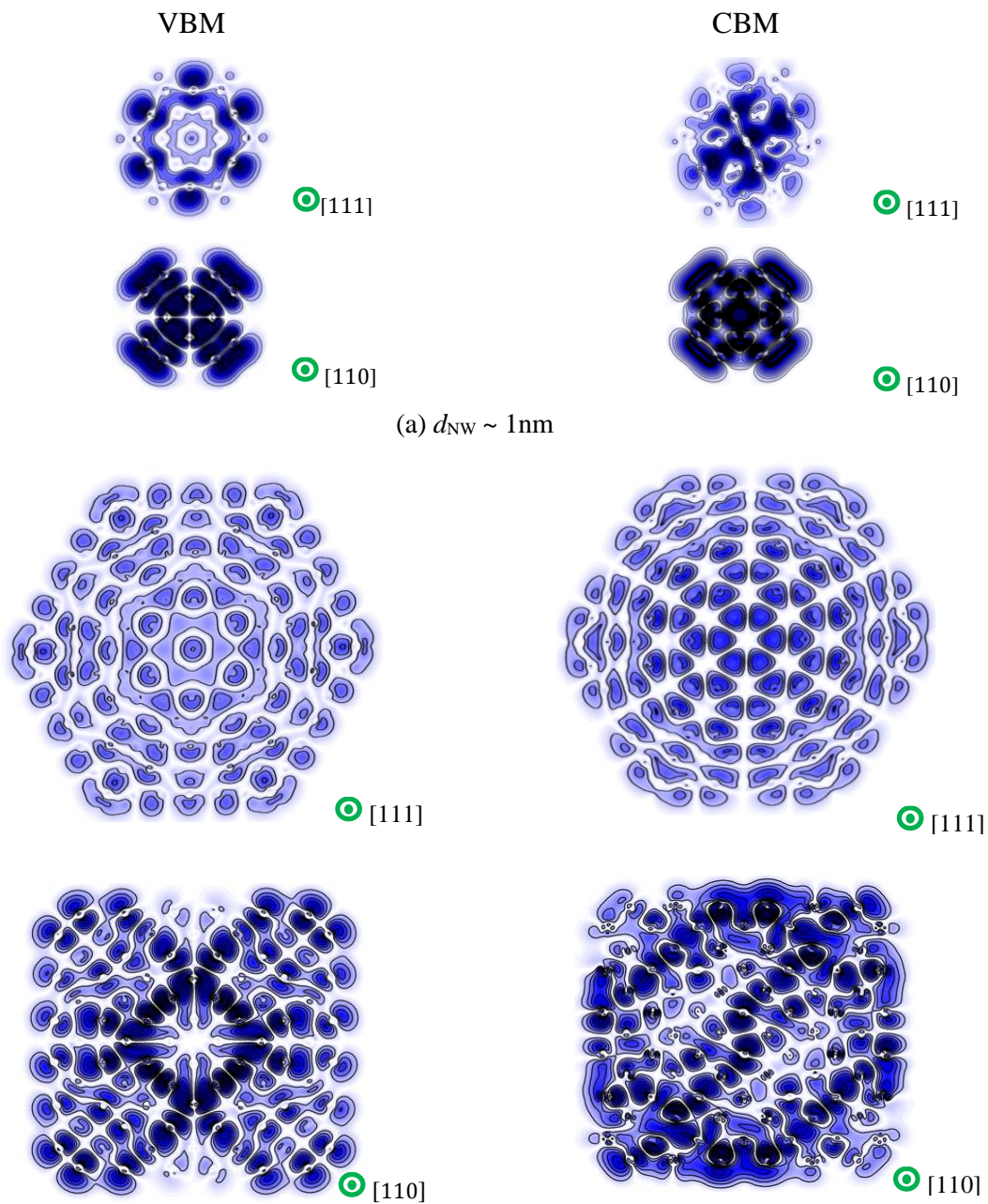

(a) $d_{NW} \sim 1$ nm

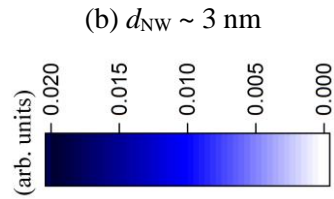

(b) $d_{NW} \sim 3$ nm

**Figure 6.** Contour plots of the charge density for the valence band minima (VBM) in the left column, and the first conduction band minima (CBM), in the right column, for (a) 1 nm and (b) 3 nm along [111] and [110] crystallographic orientations as labelled in the figure. Higher densities are indicated as blue and lower densities are indicated as white.

## 4. Summary and Conclusions

The electronic structures for [111] and [110] oriented BiNWs with critical dimensions less than 5 nm have been calculated using DFT/LDA. For the [110] oriented NWs, it is found that structural relaxations lead to a self-terminating surface as seen by a band gap emerging below the semimetal-to-semiconductor transition without energies associated with surface states being found in the energy gap. For the [111] oriented BiNWs, without surface termination the NWs remain metallic for all cross sections studied. Chemically saturating the surface bonds by hydrogen passivation removes the surface states and band gaps of relevance for room temperature electronics are found to emerge at larger diameters for the [111] oriented NWs compared to the [110] oriented NWs.

The band structure of bulk bismuth displays a 'negative band gap' with a band structure that is metallic but reflects much of the structure normally associated with semiconductors. The electronic structure in BiNWs can be as a result complex and the nature of the VBM and CBM can vary significantly with NW diameter as bands with differing effective masses are acted upon by quantum confinement. This leads to a complex behaviour for the effective masses at the VBM and CBM at differing NW diameters. For charge transport, it is important to consider the effect of charge carriers occupying partially filled bands within a voltage bias window. For BiNWs, several states with varying energy separations and effective masses will play a role in determining charge transport and determining the energy ordering of the relevant bands and their effective masses requires careful analysis.

Quasiparticle energies for the band edge states have been estimated using the *GW* approximation to provide improved estimates of band gap energies, electron affinities, ionisation potentials and work functions. As anticipated due to well-known limitations for the use of Kohn-Sham eigenstates as quasiparticle energies, the *GW* calculations predict larger band gap energies than those obtained from the DFT/LDA calculations. On the other hand, the *GW* calculations predict smaller band gap energies relative to analytical models for quantum confinement in bismuth NWs relying on the use of bulk effective masses. As the nature of the electronic states and their effective masses near the VBM and CBM edges vary substantially with respect to the bulk bismuth when patterned below 5 nm critical dimensions, analytical models should be refined to be applicable on dimensions of a few nanometers.

The variability of material parameters with NW critical dimension is not unique to bismuth, hence the variability and complexity observed in these calculations are not unlike that seen for other semiconducting nanowires. The novel properties of bulk bismuth do add a novel dimension to the electronic structure properties of BiNWs, in that the semimetal-to-semiconductor transition allows for a metallic region in a section of a BiNW with critical dimensions larger than that required for the semimetal-to-semiconductor transition to be situated next to a thinned semiconducting region with critical dimensions below that required for semimetal-to-semiconductor transition. Hence in a semimetal NW, Schottky junction formation can be achieved [30] without introducing a heterojunction or dopant atoms. BiNWs have the property that whereas materials that are semiconducting in the bulk become wide gap semiconductors in ultra thin NWs with critical dimensions of a few nanometers, on the same length scale BiNWs become semiconductors with band gaps comparable to bulk semiconductors widely in use nanoelectronics with critical dimensions greater than 7 nm.

ACKNOWLEDGMENT

This research was funded through a Science Foundation Ireland Principal Investigator award 13/IA/1956. Additional support was provided by Intel Corporation through the Intel Tyndall Collaboration. The authors would like to acknowledge the SFI/HEA Irish Centre for High-End Computing (ICHEC) for the provision of computational facilities and support.